\pdfoutput=1
\documentclass[11pt,twoside,a4paper,cmspaper,final,collab]{cms-tdr}
\def\svnVersion{5b35144}\def\svnDate{2023/09/14}\def\cmsCernNoTag{CERN-EP-2022-253}\def\cmsCernDate{\today}\def\cmsMessage{Published in the Journal of High Energy Physics as \href{http://dx.doi.org/10.1007/JHEP09(2023)057}{\doi{10.1007/JHEP09(2023)057}.}}
\begin{document}\cmsNoteHeader{B2G-21-007}

\DeclareRobustCommand{\PX}{{\HepParticle{X}{}{}}\Xspace}
\DeclareRobustCommand{\Tprime}{{\HepParticle{\PQT}{}{\prime}}\Xspace}
\DeclareRobustCommand{\Bprime}{{\HepParticle{\PQB}{}{\prime}}\Xspace}
\newcommand{\sherpa}{\textsc{sherpa}\xspace}
\newcommand{\pp}{\ensuremath{\Pp\Pp}\xspace}
\newcommand{\bW}{\ensuremath{\PQb\PW}\xspace}
\newcommand{\bq}{\ensuremath{\PQb\PQq}\xspace}
\newcommand{\tZ}{\ensuremath{\PQt\PZ}\xspace}
\newcommand{\tbW}{\ensuremath{\PQt\to\bW}\xspace}
\newcommand{\tH}{\ensuremath{\PQt\PH}\xspace}
\newcommand{\TtH}{\ensuremath{\Tprime\to \tH}\xspace}
\newcommand{\kappaT}{\ensuremath{\kappa_\PQT}\xspace}
\newcommand{\MtH}{\ensuremath{m_{\tH}}\xspace}
\newcommand{\MTprime}{\ensuremath{M_{\Tprime}}\xspace}
\newcommand{\mgg}{\ensuremath{m_{\PGg\PGg}}\xspace}
\newcommand{\qcd}{\ensuremath{\text{QCD}}\xspace}
\newcommand{\gjet}{\ensuremath{\PGg+\text{jets}}\xspace}
\newcommand{\dipho}{\ensuremath{\PGg\PGg+\text{jets}}\xspace}
\newcommand{\ttH}{\ensuremath{\ttbar\PH}\xspace}
\newcommand{\Hgg}{\ensuremath{\PH \to \PGg\PGg}\xspace}
\newcommand{\pTgg}{\ensuremath{\pt(\PGg\PGg)}\xspace}
\newcommand{\pTgone}{\ensuremath{\pt(\PGg_{1})}\xspace}
\newcommand{\pTgtwo}{\ensuremath{\pt(\PGg_{2})}\xspace}
\newcommand{\DRjg}{\ensuremath{\Delta R(j, \PGg)}\xspace}
\newcommand{\DRjl}{\ensuremath{\Delta R(j, \ell)}\xspace}
\newcommand{\DRlg}{\ensuremath{\Delta R(\ell, \PGg)}\xspace}
\newcommand{\DRlj}{\ensuremath{\Delta R(\ell, j)}\xspace}
\newcommand{\Detajg}{\ensuremath{\eta_\PGg - \eta_{j}}\xspace}
\newcommand{\Dphijg}{\ensuremath{\phi_\PGg - \phi_{j}}\xspace}

\newlength\cmsTabSkip\setlength{\cmsTabSkip}{1ex}
\providecommand{\cmsTable}[1]{\resizebox{\textwidth}{!}{#1}}
\newcolumntype{L}[1]{>{\raggedright\let\newline\\\arraybackslash\hspace{0pt}}m{#1}}
\newcolumntype{C}[1]{>{\centering\let\newline\\\arraybackslash\hspace{0pt}}m{#1}}

\cmsNoteHeader{B2G-21-007}

\title{Search for a vector-like quark \texorpdfstring{$\Tprime \to \tH$}{T' to tH} via the diphoton decay mode of the Higgs boson in proton-proton collisions at \texorpdfstring{$\sqrt{s}=13\TeV$}{sqrt(s)=13 TeV}}

\author*[cern]{The CMS Collaboration}
\date{\today}

\abstract{
    A search for the electroweak production of a vector-like quark \Tprime, decaying to a top quark and
    a Higgs boson is presented. The search is based on a sample of proton-proton collision events
    recorded at the LHC at $\sqrt{s}=13\TeV$, corresponding to an integrated luminosity of
    138\fbinv. This is the first \Tprime search that exploits the Higgs boson decay to a pair of photons.
    For narrow isospin singlet \Tprime states with masses up to 1.1\TeV, the excellent diphoton invariant
    mass resolution of 1--2\% results in an increased sensitivity compared to previous searches based on the same production mechanism.
    The electroweak production of a \Tprime quark with mass up to 960\GeV is excluded at 95\% confidence level,
    assuming a coupling strength $\kappaT = 0.25$ and a relative decay width $\Gamma/\MTprime < 5$\%.}

\hypersetup{
    pdfauthor={CMS Collaboration},
    pdftitle={Search for a vector-like quark T' to tH via the diphoton decay mode of the Higgs boson in proton-proton collisions at sqrt(s)=13 TeV},
    pdfsubject={CMS},
    pdfkeywords={CMS, VLQ, Higgs, diphoton}}

\maketitle
\section{Introduction}
The Higgs boson (\PH) was discovered by the ATLAS and
CMS Collaborations in 2012~\cite{the_atlas_collaboration_observation_2012,
    the_cms_collaboration_observation_2012, cms_collaboration_observation_2013}. With this discovery,
the standard model (SM) is now in principle complete as a low-energy effective theory, describing
all known fundamental particles and their interactions. However, the stability of the Higgs boson
mass at the electroweak (EW) scale remains unexplained: as the SM is extrapolated to high energies,
quantum loop corrections to the Higgs boson self-energy quadratically diverge~\cite{susskind_dynamics_1979}.
Various theories beyond the SM predict additional particles that can affect these quantum
corrections to the Higgs boson mass. One such new particle is a vector-like quark (VLQ).

The VLQs are hypothetical spin-$\frac{1}{2}$, colored particles whose left- and right-handed
components transform in the same way under the SM gauge group. Therefore, unlike the chiral
quarks in the SM, their masses are not generated by a Yukawa coupling to the Higgs boson and have
a lower contribution to the production cross section of the Higgs boson. A variety of new physics
models, such as composite Higgs models~\cite{contino_light_2007, contino_warpedcomposite_2007,
    kaplan_flavor_1991, dugan_anatomy_1985, blasi_softened_2019}, little Higgs
models~\cite{perelstein_top_2004, matsedonskyi_light_2013, schmaltz_little_2005}, and models with a
warped extra dimension~\cite{randall_large_1999}, incorporate VLQs which provide solutions to the above mentioned
Higgs boson mass stability problem. In minimal models, the VLQs may only exist as electroweak singlets,
denoted as \Tprime and \Bprime, and doublets, carrying respective electric charges
of 2/3$e$ and -1/3$e$; further doublets and triplets may also exist where the VLQs can have
exotic charges.

\begin{figure}[htbp!]
    \centering
    \includegraphics[width=0.5\textwidth]{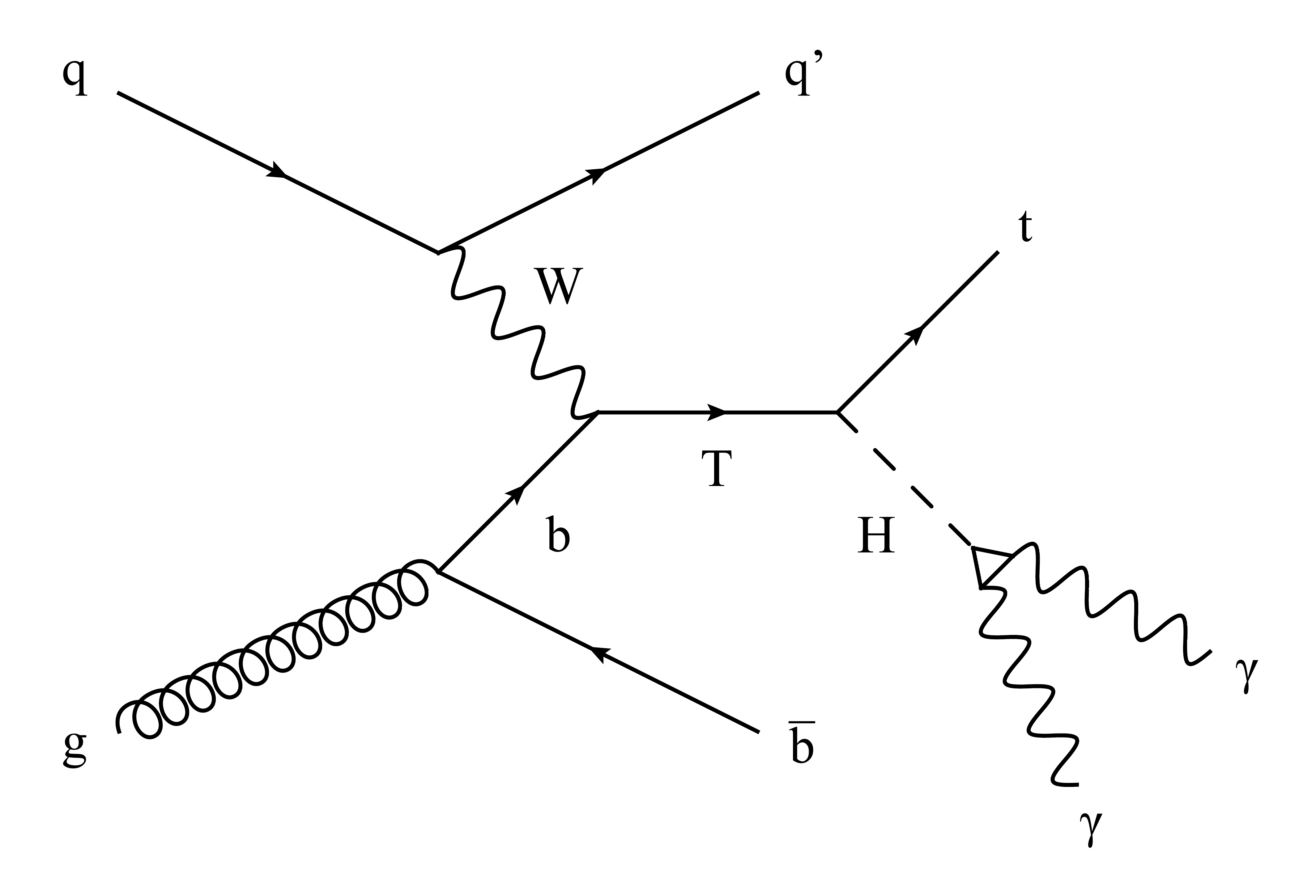}
    \caption{Leading-order Feynman diagram for single \Tprime production in Wb fusion and its subsequent decay into tH (\Hgg)}
    \label{fig:Fyn}
\end{figure}

Here, we present a search for the production of a vector-like top quark
partner \Tprime at the LHC.
These could either be pair-produced by the strong interaction or singly-produced by the electroweak one.
For pair production through the strong interaction,
the available parton level center-of-mass energy is shared between the two heavy particles. In contrast,
in the EW production of a single \Tprime, a larger kinematic phase space is available and heavier
masses can, in principle, be probed. The \Tprime quark can couple to SM quarks and charged or neutral bosons, resulting
in decays to \bW, \tZ, and \tH channels. For the EW production of an isospin singlet \Tprime VLQ, considered
in this paper, the \Tprime branching fractions ($\mathcal{B}$) are assumed to be 50, 25, and 25\%, respectively,
for \bW, \tZ, and \tH decays~\cite{VLQ_Handbook}. The leading order Feynman diagram for the electroweak
\Tprime production along with the \Hgg decay is shown in Fig.~\ref{fig:Fyn}. The EW production cross section depends explicitly
on the couplings of the
VLQ to third-generation quarks~\cite{VLQ_Handbook, Deandrea:2021vje}, henceforth refered as \kappaT. Here, the coupling \kappaT can
significantly change based on the choice of the VLQ mass and width. In this study, the VLQ
production and decay are parametrized using the narrow width approximation (NWA)~\cite{NWAreference},
where the \Tprime natural width ($\Gamma$) is approximately 1\% relative to its mass (\MTprime).
The NWA is valid up to $\Gamma/\MTprime \approx 10$--15\%, beyond which the large width of \Tprime and its
interference with the SM background become important~\cite{Deandrea:2021vje, Banerjee:2021hfo}.
However, the sensitivity of this analysis extends up to $\Gamma/\MTprime\approx 5\%$, which roughly
corresponds to the experimental resolution of \MTprime.

This search is based on the pp collision data recorded with the CMS detector during LHC
operations from 2016--2018, corresponding to an integrated luminosity of 138\fbinv.
The study focuses on the EW production of \Tprime in proton-proton (\pp) collisions at
$\sqrt{s} = 13\TeV$, $\pp \to \Tprime\bq$, followed by the decay of $\TtH$, where the Higgs
boson decays into a pair of photons ($\Hgg$). The leptonic and hadronic decay modes of the top
quark are treated separately to achieve the best possible search sensitivity. It is the first \Tprime search by the LHC experiments in $\Hgg$ channel.
The ATLAS and CMS Collaborations have previously performed searches for strong and EW
production of VLQs at $\sqrt{s} =13\TeV$~\cite{ATLAS-Aaboud_2018, cms_collaboration_search_2019,
    cms_collaboration_search_2020,atlas_collaboration_2022, CMS:2022fck, ATLAS:2022tla}.
The most recent results on pair production \cite{CMS:2022fck, ATLAS:2022tla} exclude
\Tprime masses below 1.48\TeV at 95\% confidence level (\CL), assuming branching fractions
of 50, 25 and 25\% for \bW, \tZ, and \tH decays, respectively.

These previous searches used the reconstructed \Tprime invariant mass or transverse mass
as the main observable. The present analysis exploits the excellent resolution of 1--2\%
for the reconstructed H mass in the diphoton decay channel to search for a signal characterized by
a peak at the H mass above the falling diphoton mass (\mgg) continuum.
The statistical methodologies and the non-Higgs background determination techniques are identical
to the ones used in the SM $\Hgg$ measurements~\cite{CMS:2021kom}.

\section{The CMS detector}
The CMS apparatus~\cite{Chatrchyan:2008zzk} is a multipurpose, nearly hermetic detector, designed to
trigger on~\cite{CMS:2020cmk,CMS:2016ngn} and identify electrons, muons, photons, and charged
and neutral hadrons~\cite{ele_pho_reco,muon_reco,CMS:2014pgm}. A global ``particle-flow'' (PF)
algorithm~\cite{Sirunyan:2017ulk} aims to reconstruct all individual particles in an event, combining
information provided by the all-silicon inner tracker, by the crystal electromagnetic calorimeters (ECAL), and
brass-scintillator hadron calorimeters, operating inside a 3.8\unit{T} superconducting solenoid,
with data from the gas-ionization muon detectors embedded in the flux-return yoke outside the
solenoid.
Events of interest are selected using a two-tiered trigger system. The first level of the trigger
system, composed of special hardware processors, uses information from the calorimeters and muon
detectors to select the most interesting events in a time interval of less than 4\mus. The second
level, known as the high-level trigger, consists of a farm of processors that further decreases
the event rate from around 100\unit{kHz} to less than 1\unit{kHz}~\cite{CMS:2016ngn}, running a
version of the full event reconstruction software optimized for fast processing.
The primary vertex is taken to be the vertex corresponding to the hardest scattering
in the event, evaluated using tracking information alone, as described in Section~9.4.1 of
Ref.~\cite{CMS-TDR-15-02}. The 2017 and 2018 data sets benefit from the upgrade of the pixel
tracking detector in the winter of 2016--2017~\cite{CMS:2012sda}, improving the acceptance,
redundancy, and resolution. The updated tracker greatly enhances the performance of \PQb jet
identification~\cite{CMS-DP-2018-033}, which is essential to analysis and online event selection.

\section{Simulated samples}
The data analysis strategy has been optimized using Monte Carlo (MC) simulation, where specific
\pp event generators and a \GEANTfour-based detector simulation~\cite{geantreference} have been
utilized. The signal process, $\pp \to \Tprime (\to \tH)\bq$~\cite{Deandrea:2021vje}, is generated
to leading order and with $\Gamma/\MTprime \approx 1\%$.
Samples of events are generated with \textsc{MadGraph5\_aMC@NLO}~2.3.3~\cite{Alwall_2014, Artoisenet_2013}
at ten \MTprime points, from 600 to 1200\GeV, using NNPDF3.1~\cite{NNPDF3.1} as the parton
distribution function (PDF) set. The Higgs boson and
top quark masses are set to 125.0 and 172.5\GeV, respectively. The SM background processes
contributing to the \mgg spectrum are categorized in two types: SM Higgs boson (SMH) background and SM nonresonant
background (NRB). The \MGvATNLO generator has been utilized to simulate the SM Higgs boson
production, including gluon fusion (ggH)~\cite{bagnaschi_higgs_2012}, vector-boson
fusion (VBF)~\cite{nason_nlo_2010}, production in association with top quarks (\tH, \ttH)~\cite{hartanto_higgs_2015},
or with a vector boson (VH)~\cite{luisoni_hwhz_2013} at next-to-leading order in quantum chromodynamics
(\qcd). The total cross sections and branching fractions, as recommended by the LHC Higgs boson
cross section working group~\cite{cern_higgs_cross_section_2017}, have been adopted for the SMH
production processes. The background processes, $\PQt+\PX$, $\ttbar+\PX$, $\PW\PGg$, and
$\PZ\PGg$, are simulated with \MGvATNLO, whereas diboson events are produced at the leading order
with \PYTHIA8.205~\cite{Sj_strand_2015}. The nonresonant diphoton samples are simulated with
\sherpa~2.2.4~\cite{Bothmann_2019} which includes tree-level processes with up to three additional
partons, as well as box diagrams. For all MC samples generated with \MGvATNLO, the parton showering and hadronization have
been implemented via \PYTHIA, with the underlying event tune CUETP8M1~\cite{Khachatryan_2016} and
CP5~\cite{Sirunyan_2020_CP5} respectively for the 2016 and the 2017--2018 datasets. The nonresonant
background samples are used to train a multivariate analysis (MVA) discriminant, while the corresponding
yields are extracted using sidebands, defined by $\mgg < 115$ or $\mgg > 135\GeV$, in data.

\section{Event selection}
The online event selection of the diphoton triggers requires at least two photons with asymmetric
conditions on the photon transverse momenta: $\pTgone > 30\GeV$ and $\pTgtwo > 18$ or $>$22\GeV,
depending on the data taking period. Moreover, requirements~\cite{CMS:2021kom} on the
isolation in the calorimeter and on the shape of the electromagnetic shower are imposed. The \mgg
is required to be above 90\GeV, assuming both photons originate from the primary vertex.

For efficient selection of photons associated with the primary vertex, a separate MVA based
on the isolation and photon shower shape variables, called the photon `ID MVA'~\cite{CMS:2021kom},
is used in the offline event selection.
Events are selected with at least two ID MVA selected photons within the ECAL and the tracker
fiducial region (pseudorapidity $\abs{\eta} < 2.5$, but excluding the ECAL barrel-endcap
transition region, $1.44 < \abs{\eta} < 1.57$). The photon pairs are further
required to satisfy the criteria: {$100 < \mgg < 180\GeV$}, {$\pTgone/\mgg > 1/3$}, and
{$\pTgtwo/\mgg > 1/4$}; in the case of multiple diphoton pairs, the one with highest
$\pTgg$ is chosen~\cite{CMS:2021kom}.

Reconstructed particles are used to form jets using the anti-\kt algorithm with a distance
parameter of 0.4~\cite{AK4, Cacciari:2011ma}, and to estimate the missing transverse momentum
(\ptmiss)~\cite{CMS:2018jrd,CMS:2016lmd,METperformance}.
This search uses energy-corrected jet candidates with $\pt > 25\GeV$ and $\abs{\eta} < 4.5$, with
stringent requirements imposed to remove spurious jets~\cite{JINST6}. Jets must be separated
from photons and leptons in the event, with
$\DRjg\equiv \sqrt{\smash[b]{(\Detajg)^2 +(\Dphijg)^2}}  > 0.4 $ and $\DRjl > 0.4$, where $\phi$ is the
azimuthal angle in radians. For the identification (tagging) of jets from \PQb quark hadronization
and decay (\PQb jets), the deep neural network based \textsc{DeepCSV} algorithm~\cite{CMS:2017wtu}
is applied, for jets with $\abs{\eta} < 2.5$. This search utilizes the \PQb-tagged jets with
\textsc{DeepCSV} scores exceeding a minimum value corresponding to a misidentification probability
of 10\% for light quark and gluon jets, and to an identification efficiency for \PQb jets of
75--90\% depending on the jet \pt~\cite{CMS:2017wtu}.

Isolated leptons with $\pt > 10\GeV$ and within appropriate fiducial volumes
($\abs{\eta} < 1.44$ or $1.57 < \abs{\eta} < 2.40$, and $\abs{\eta} < 2.40$, respectively, for
electrons and muons) are considered in this search. These leptons are further required to be
separated from any photon or jet considered in the analysis: $\DRlg > 0.4$ and $\DRlj > 0.4$.
Neutrinos are accounted for through the reconstruction of \ptmiss.
The \ptvecmiss vector is computed as the negative vector \pt sum of all the PF candidates in an event~\cite{METperformance}.
The \ptvecmiss is modified to account for corrections to the energy scale of the reconstructed jets in the event.

As mentioned above, this search categorizes events based on the leptonic or hadronic decays
of the top quark. Events containing a pair of photons and at least one electron or muon
are defined as the leptonic category; those with a pair of photons and no lepton form the hadronic
category. To target the \tbW decay, at least one \PQb-tagged jet is required in the leptonic channel,
and three jets, of which at least one is \PQb tagged, are required in the hadronic channel.
Events with two leptons from the Drell--Yan processes contribute to the background in
the leptonic category and are rejected by requiring $\abs{m_{\Pe\Pe/\Pgm\Pgm} - M_\PZ} > 5\GeV$.
In the leptonic category, the \qcd, \gjet, and \dipho processes
constitute 25\% of the total background yield. In the hadronic category, however, these background
contributions amount to 95\% of the total. 
Owing to imperfect MC modeling of the \gjet and multi-jets processes, those backgrounds
are estimated from data in a phase space region defined by inverting the selection on the
photon ID MVA, which consists of a sample of events with a significant contribution from
jets with at least one of them misidentified as a photon. For each of these events, the photon
ID MVA value of the photon candidate failing the ID MVA selection is substituted with a
value generated from the MC distribution of photon ID MVA for misidentified jets passing
the photon ID requirement. This is the same procedure used in the previous \Hgg related
studies \cite{CMS:2020cga, CMS:2021hug}. Since the simulation of the \dipho process is
performed only at leading order~\cite{CMS:2021kom}, additional normalization factors
accounting for higher orders are derived from a simultaneous template fit to the photon
ID MVAs. The normalization factors are then applied to the control samples in data
(\gjet and multi-jets) and the \dipho samples when using those samples to provide the
background in the MVA training. However, these normalization factors have no role in
creating the final background model of the analysis.

\section{Discrimination between signal and background}
\begin{figure}[htbp!]
    \centering
    \includegraphics[width=0.49\textwidth]{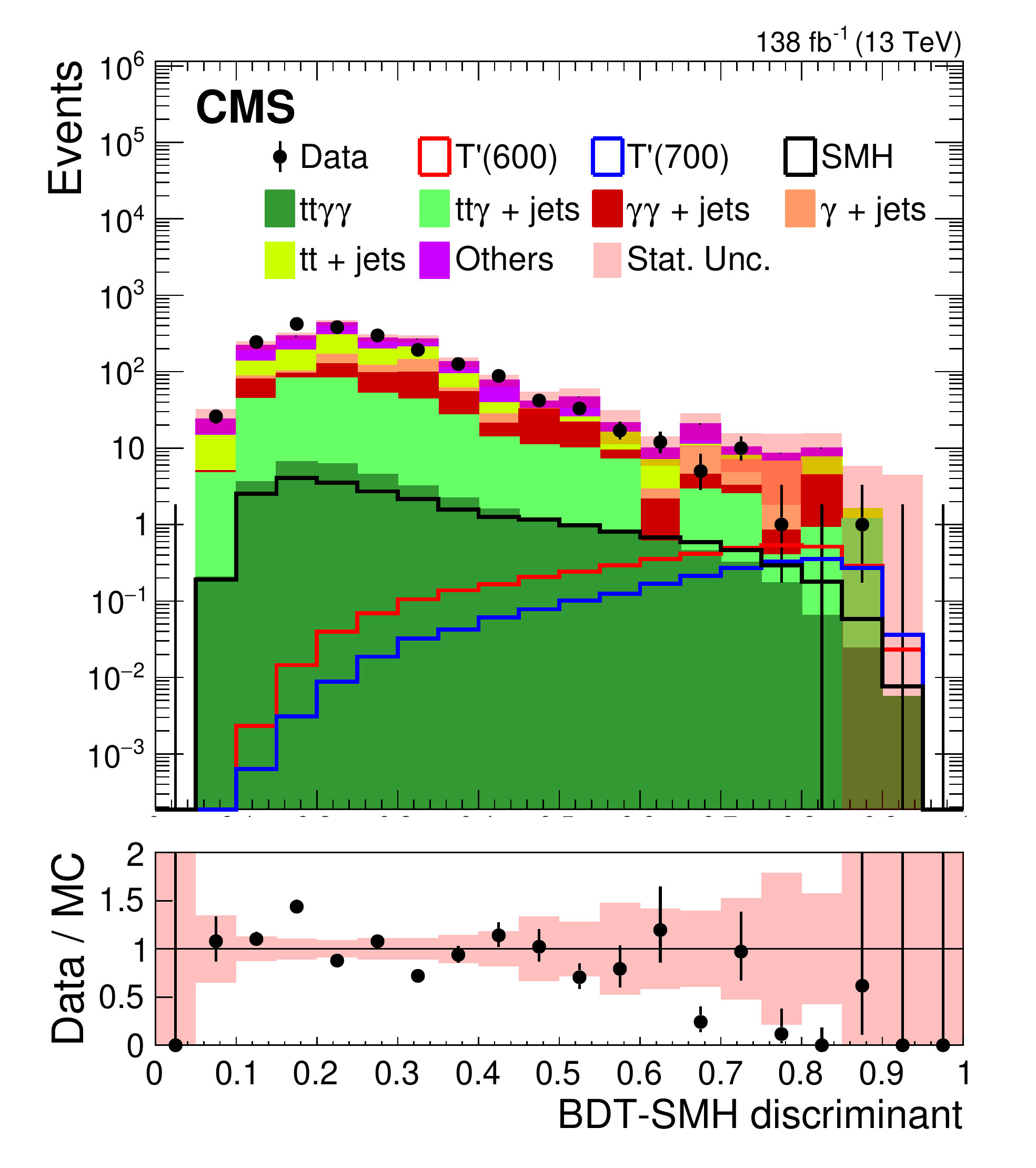}
    \includegraphics[width=0.49\textwidth]{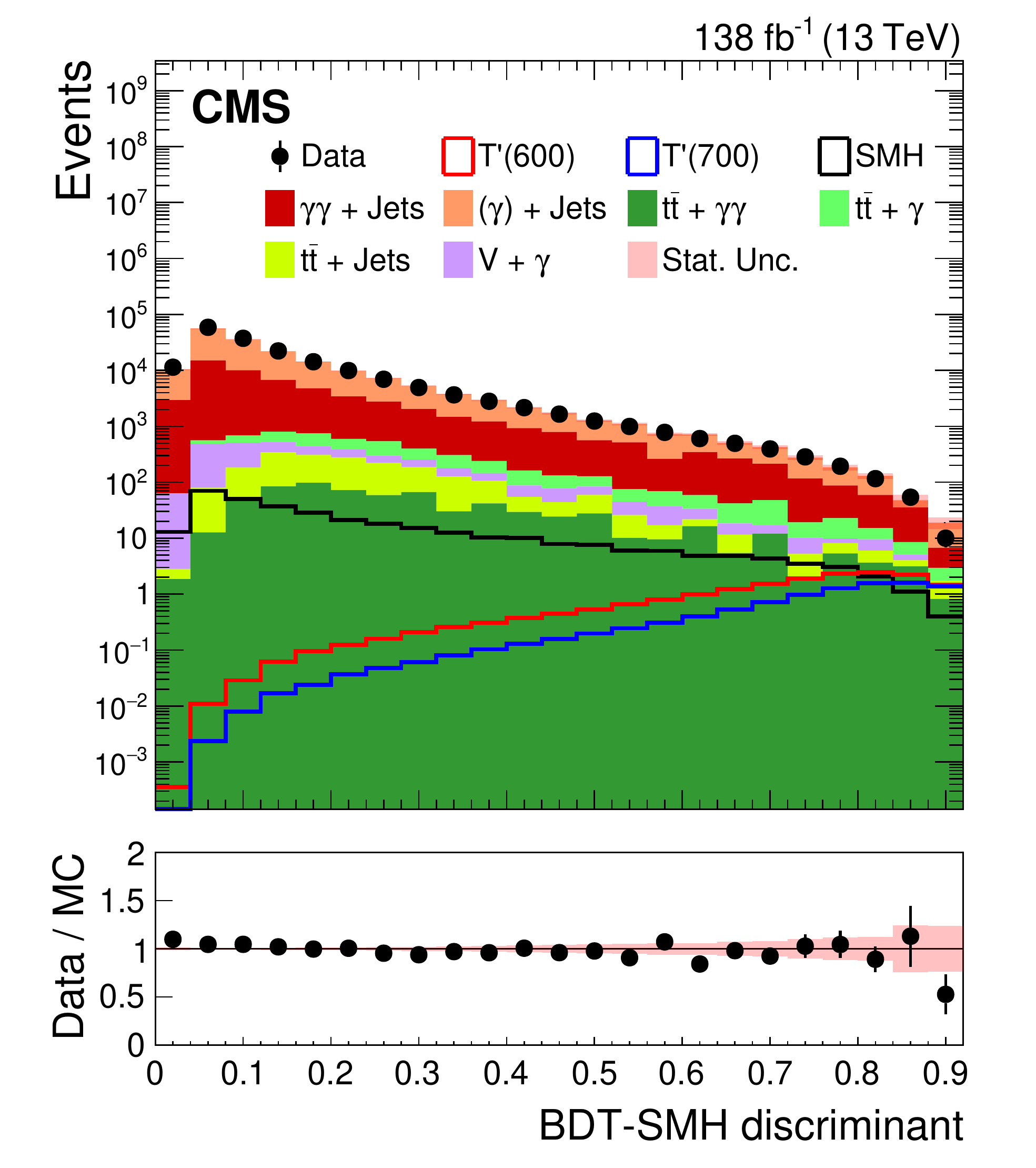}
    \includegraphics[width=0.49\textwidth]{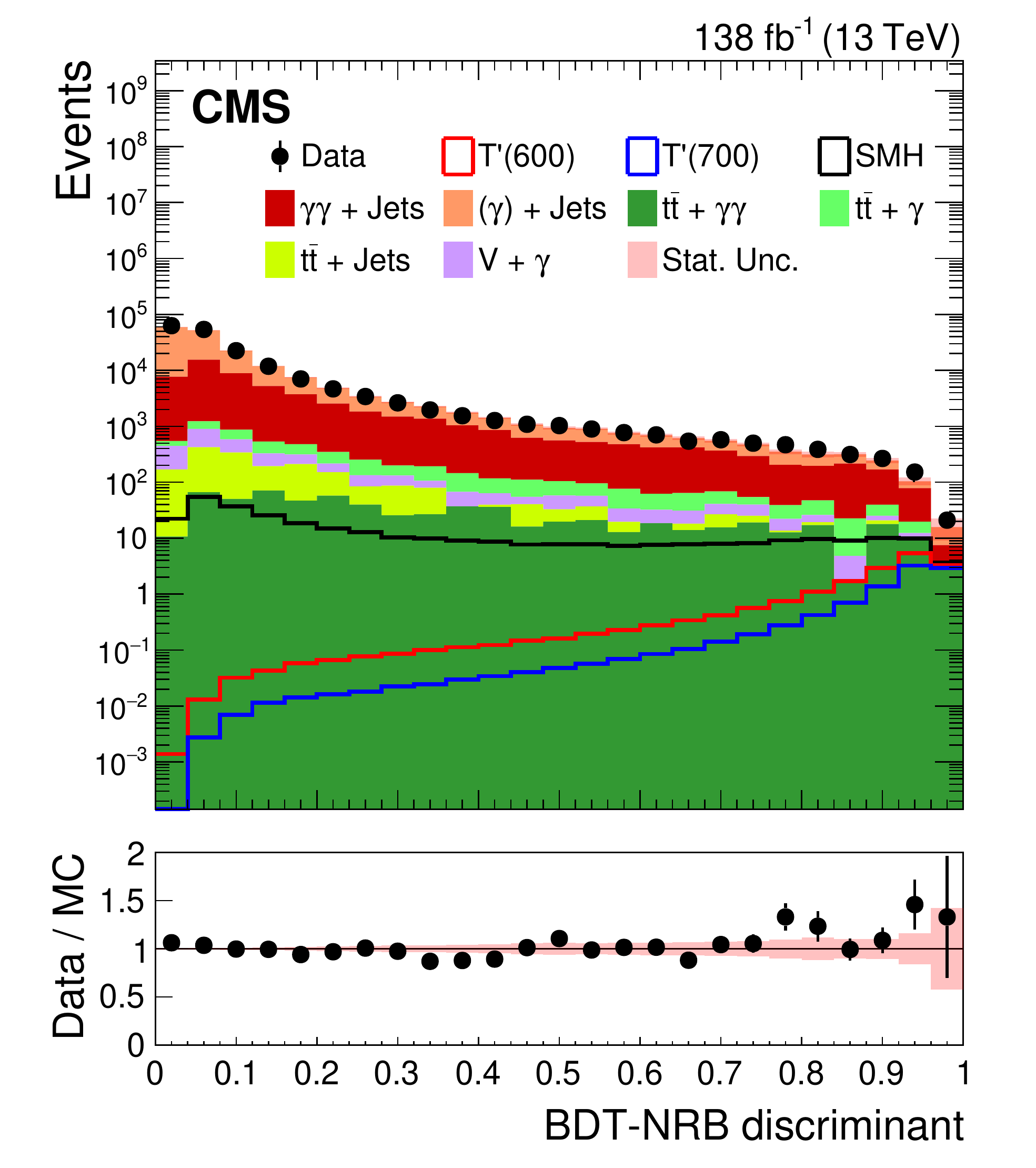}
    \caption{The BDT output distributions for data, backgrounds and signal events in the leptonic and the hadronic categories:
        leptonic BDT trained against the SM Higgs boson backgrounds (upper left),
        hadronic BDT trained against the SM Higgs boson backgrounds (upper right),
        and hadronic BDT trained against the nonresonant backgrounds processes (lower).
        For the leptonic category, MC-estimated nonresonant backgrounds are normalized to the number of observed data events.
        For the hadronic category, the control samples in data have been adapted for \gjet backgrounds, while all other
        MC samples are normalized to an integrated luminosity of 138\fbinv. The shaded region
		in the lower panels represents the statistical uncertainties of the different samples.
    }
    \label{fig:bdtg_scores}
\end{figure}

At this level of selection, \ttH with \Hgg is the dominant background among the SMH production
processes for both categories, since it also leads to a peak in the \mgg spectrum at the Higgs boson mass.
The \mgg spectrum from \Tprime signal also peaks at $M_{\PH}$ due to \TtH decay.
To separate the overlapping \Tprime signal from the SMH background processes,
MVA discriminants based on boosted decision trees (BDTs) are implemented~\cite{hoecker_tmva_2009}
separately for each category (BDT-SMH). Furthermore, an additional BDT (BDT-NRB) is trained to
suppress the sizeable nonresonant background contributions in the hadronic category.
In the leptonic channel, discrimination against the background exploits characteristic features of the kinematic
properties of each of the objects contributing to the signal (two prompt photons, \PQb jet,
lepton and neutrino) and on energy and momentum conservation.
Similarly for the hadronic channel, the input features of both BDT-SMH and BDT-NRB include
the kinematic properties of the physics objects: photons, jets, diphotons, reconstructed
top quark candidates, jet multiplicities, \ptmiss, \PQb-tagging scores of jets from the DeepCSV algorithm, and the
output of the photon ID MVA for both photons.
In order to prevent a possible correlation between \mgg and the BDT score, the ratios \pTgone/\mgg, \pTgtwo/\mgg, and \pTgg/\mgg
are provided as input to the BDT training, rather than using \mgg directly. The linear correlation 
between the BDT score and \mgg is found to be less than 5\% in all samples.
As the kinematic distributions of the signal vary among different \MTprime values, separate BDTs
have been used in three \MTprime ranges: 600--700, 700--1000, and 1000--1200\GeV.
In the leptonic category, the trained BDT-SMH yields a signal efficiency of more than 75\%,
for a background efficiency of 10\%. Similarly, in the hadronic category, the trained BDT-SMH
and BDT-NRB yield a signal efficiency of more than 96\% and 98\%, respectively, for a background
efficiency of 10\%. Furthermore, each BDT output distribution of the hadronic and leptonic
categories in the \mgg sideband region shows an overall statistical agreement between data and
simulation with a $\chi^2$ p-value of 5\%. The fluctuations observed in some of the distributions
are associated with a limited simulated sample size.
Figure~\ref{fig:bdtg_scores} shows the distribution of the well-separated BDT output scores
for signal and background processes when the training is performed over the \Tprime samples
having masses between 600 and 700 GeV.

The kinematic properties of each Higgs boson candidate are always reconstructed from the momenta of the
photons. However, the kinematic distributions of the top quark candidates are reconstructed
differently for leptonic and hadronic channels; the leptonic channel uses constraints on the
event \pt conservation and the \PW boson mass~\cite{CMS:2018hfk}; the hadronic channel uses the
minimum $\chi^2$ method~\cite{CMS:2017bhz} to choose the correct combination of jets. The reconstructed
top quark mass is also
used as one of the input variables for the BDTs in the hadronic channel. The candidate \Tprime mass,
$\MtH$, is reconstructed by combining the four-momenta of the top quark and the H candidates,
with an overall experimental resolution of 5--7\% in both leptonic and hadronic
channels. To maximize the selection efficiency, the events considered by each optimized BDT are required
to fall within a broad window in \MtH that extends beyond the range of \MTprime for which that BDT was trained with.

The primary experimental observable for this search is the diphoton invariant mass, \mgg.
Higgs bosons from both SM processes and \Tprime decay are expected
to peak on a smoothly falling \mgg distribution in the range $100 < \mgg < 180\GeV$. The selection
criteria on the BDT discriminants for the three different \MTprime ranges, labelled I, II, III
in Table.~\ref{tab:sequential_cuts}, have been optimized to maximize signal sensitivity with respect
to the background, separately for the leptonic and hadronic categories.
For statistically robust modelling of the NRB from data, the optimization requires at least eight events
in the signal sideband regions of \mgg.
The $\MtH$ window criterion is the same
between leptonic and hadronic channels. The complete list of selection criteria on BDT score and the \MtH window are provided in
Table.~\ref{tab:sequential_cuts}.
The expected yield of a \Tprime for $\kappaT$ fixed at 0.2, nonresonant background, and the SM Higgs boson background processes
within the signal region (SR), $\mgg \in [115, 135]$\GeV, for each signal window are shown in the
Table.~\ref{tab:yields_SR} together with the observed number of events in the SR.

\begin{table} [htbp!]
    \renewcommand{\arraystretch}{1.2}
    \centering
    \topcaption{
        Signal selection criteria for the three BDTs and \MtH windows.
    }
    \begin{tabular}{cC{0.12\textwidth}C{0.13\textwidth}C{0.13\textwidth}}
        BDT                   & I           & II            &   III\\
        \hline
        $\MTprime (\GeVns{})$ & [600, 700] & [700, 1000] & [1000, 1200] \\
        \hline
        \multicolumn{4}{c}{Hadronic analysis}\\
        BDT-NRB score & $>$ 0.94 & $>$0.96 & $>$0.95 \\
        BDT-SMH score & $>$0.80 & $>$0.80 & $>$0.80 \\
        $\MtH$ window (\GeVns{}) & [480, 800] & [550, 1150] & [650, 1600] \\
        \hline
        \multicolumn{4}{c}{Leptonic analysis}\\
        BDT score & $>$0.60 & $>$0.40 & $>$0.40\\
        $\MtH$ window (\GeVns{})  & [480, 800] & [550, 1150] & [650, 1600] \\
        \hline
    \end{tabular}
    \label{tab:sequential_cuts}
\end{table}

\begin{table} [htbp!]
    \renewcommand{\arraystretch}{1.2}
    \centering
    \topcaption{
        The expected yields of different processes in each signal window for events with a
        \Tprime with mass in the range $\MTprime\in [600, 1200]$\GeV, and the observed number of events in the signal region $\mgg \in [115, 135]$.
        Here, the yields for the \Tprime are for $\kappaT$ fixed at 0.2.
    }\cmsTable{
    \begin{tabular}{c  c c@{\hspace*{5pt}} c c c c c@{\hspace*{5pt}} c c c c}
         &&&
        \multicolumn{4}{c}{Leptonic yield}&&
        \multicolumn{4}{c}{Hadronic yield}\\
       \cline{1-2} \cline{4-7}\cline{9-12}
        \multirow{2}{*}{BDT} 			&\multirow{2}{*}{\MTprime (\GeVns)}&&\multirow{2}{*}{\Tprime} & Nonres. & SM \PH &\multirow{2}{*}{Obs.} &&\multirow{2}{*}{\Tprime} & Nonres. & SM \PH&\multirow{2}{*}{Obs.}\\
        & 			&&  & bkgd. 	& bkgd.  &  	&&	  & bkgd. 	& bkgd. & \\
        \hline
        \multirow{5}{*}{I} & 600 && 1.7 &
        \multirow{5}{*}{$11.0\pm 9.0$} &
        \multirow{5}{*}{$1.3\pm0.1$} & \multirow{5}{*}{1} && 3.2 & \multirow{5}{*}{$1.6\pm0.9$} & \multirow{5}{*}{1.8 $\pm$ 0.1} & \multirow{5}{*}{4}\\
        & 625 && 1.7 & & & && 3.5 & & &\\
        & 650 && 1.6 & & & && 3.6 & & &\\
        & 675 && 1.6 & & & && 3.7 & & &\\
        & 700 && 1.5 & & & && 3.6 & & &\\[\cmsTabSkip]
        \multirow{3}{*}{II} &  800 && 1.6 & \multirow{3}{*}{$19.0 \pm 14.4$} &
        \multirow{3}{*}{$2.3  \pm  0.1$} & \multirow{3}{*}{16} && 2.9 & \multirow{3}{*}{$7.3 \pm 4.0$} & \multirow{3}{*}{$2.0\pm 0.1$} & \multirow{3}{*}{6}\\
        & 900 && 1.2 & & & && 3.0 & & &\\
        & 1000 && 0.8 & & & && 2.5 & & &\\[\cmsTabSkip]
        \multirow{2}{*}{III} & 1100 && 0.7 & \multirow{2}{*}{$14.4 \pm 13.7$} & \multirow{2}{*}{$1.4  \pm  0.1$} & \multirow{2}{*}{10} && 2.3 & \multirow{2}{*}{$9.0\pm 5.3$} & \multirow{2}{*}{$2.4\pm 0.2$} & \multirow{2}{*}{7}\\
        & 1200 && 0.5 & & & && 1.8 & & &\\
        \hline
    \end{tabular}}
    \label{tab:yields_SR}
\end{table}

\section{Signal and background modeling}
Models of the signal and SMH background processes are obtained by fitting the
\mgg distributions in simulation with a sum of at most five Gaussian functions,
separately for each category.

The models used to describe the nonresonant background processes are extracted
from the observed \mgg spectrum in the region $\mgg \in [100, 180]\GeV$
using a discrete profiling method~\cite{envelope}.
This technique accounts for the systematic uncertainty in the background estimate associated
with choosing a particular analytic function to describe the \mgg spectrum.
The chosen functions are from a list of families of functions:
exponentials, power laws, polynomials, and Laurent series~\cite{envelope}.
However, the degrees of freedom for these functions are decided in each case using
a detailed $\mathcal{F}$-test~\cite{fisher_1922} with a loose requirement on the goodness-of-fit.

\section{Systematic uncertainties}
Systematic uncertainties that modify the \mgg distributions are incorporated
in the signal shape and normalization as nuisance parameters.
The dominant experimental uncertainties affecting the event yields and signal
shape are those associated with:
the integrated luminosity~\cite{CMS-LUM-17-003,CMS-PAS-LUM-17-004,CMS-PAS-LUM-18-002},
the photon identification MVA ID score, the jet energy scale and resolution, the trigger
efficiency, the dependence of the selection efficiency on photon shower-shape
variables, the estimation of \ptmiss, the corrections to the photon energy scale and
resolution, and the \PQb jet identification efficiency.

In addition, theoretical uncertainties arise from the variations of the \qcd
renormalization and factorization scales involved in the cross section computation
of the SM processes. The uncertainties that account for the limited knowledge of the
PDFs and the \Hgg branching fractions are also included.
The impact of each systematic uncertainty on the final signal strength ($\mu = \sigma/\sigma_{\mathrm{th}}$) is less than 5\%, and the results are limited only by the statistical uncertainties.

\begin{figure}[htbp!]
    \centering
    \includegraphics[width=0.49\textwidth]{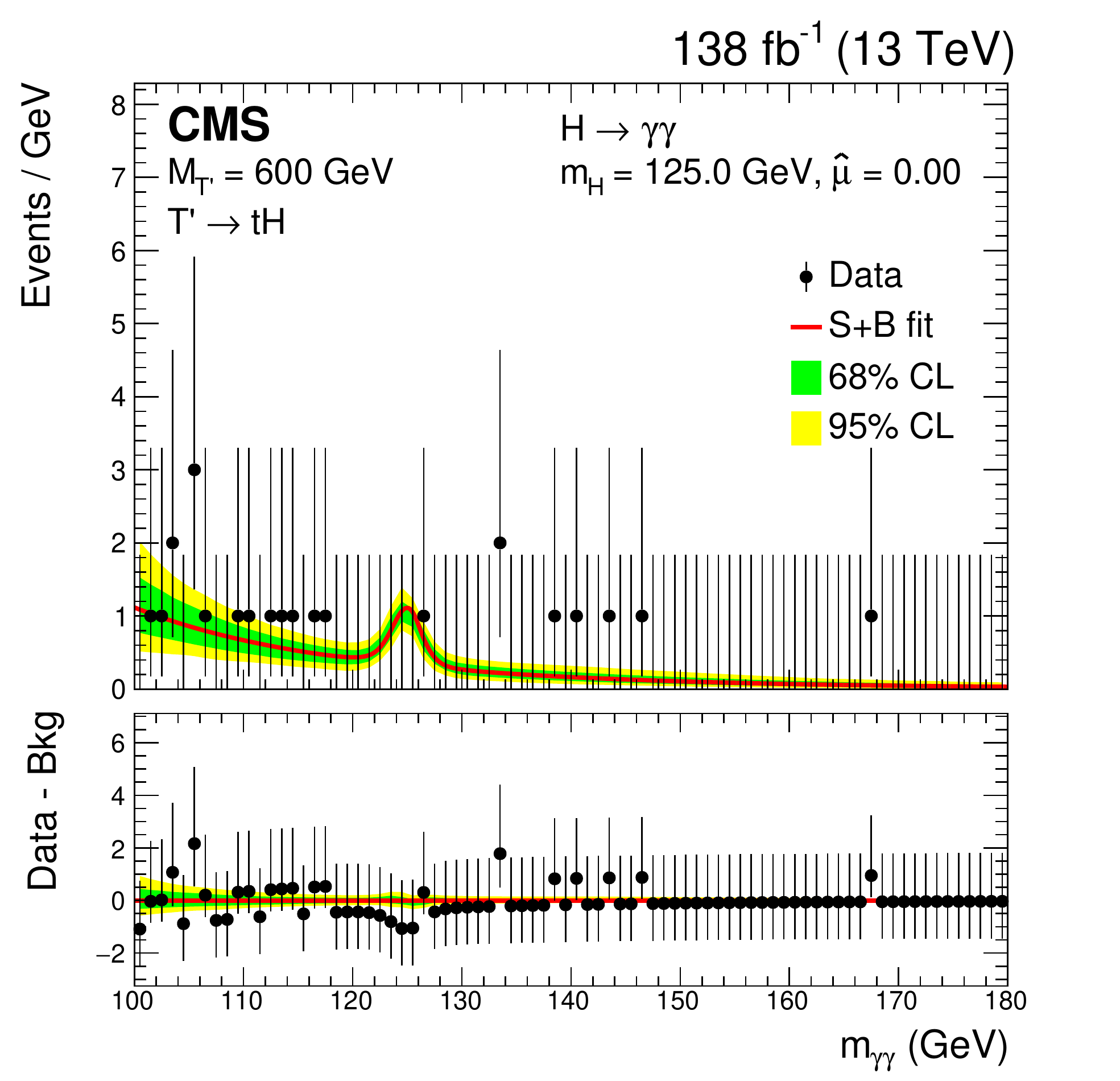}
    \includegraphics[width=0.49\textwidth]{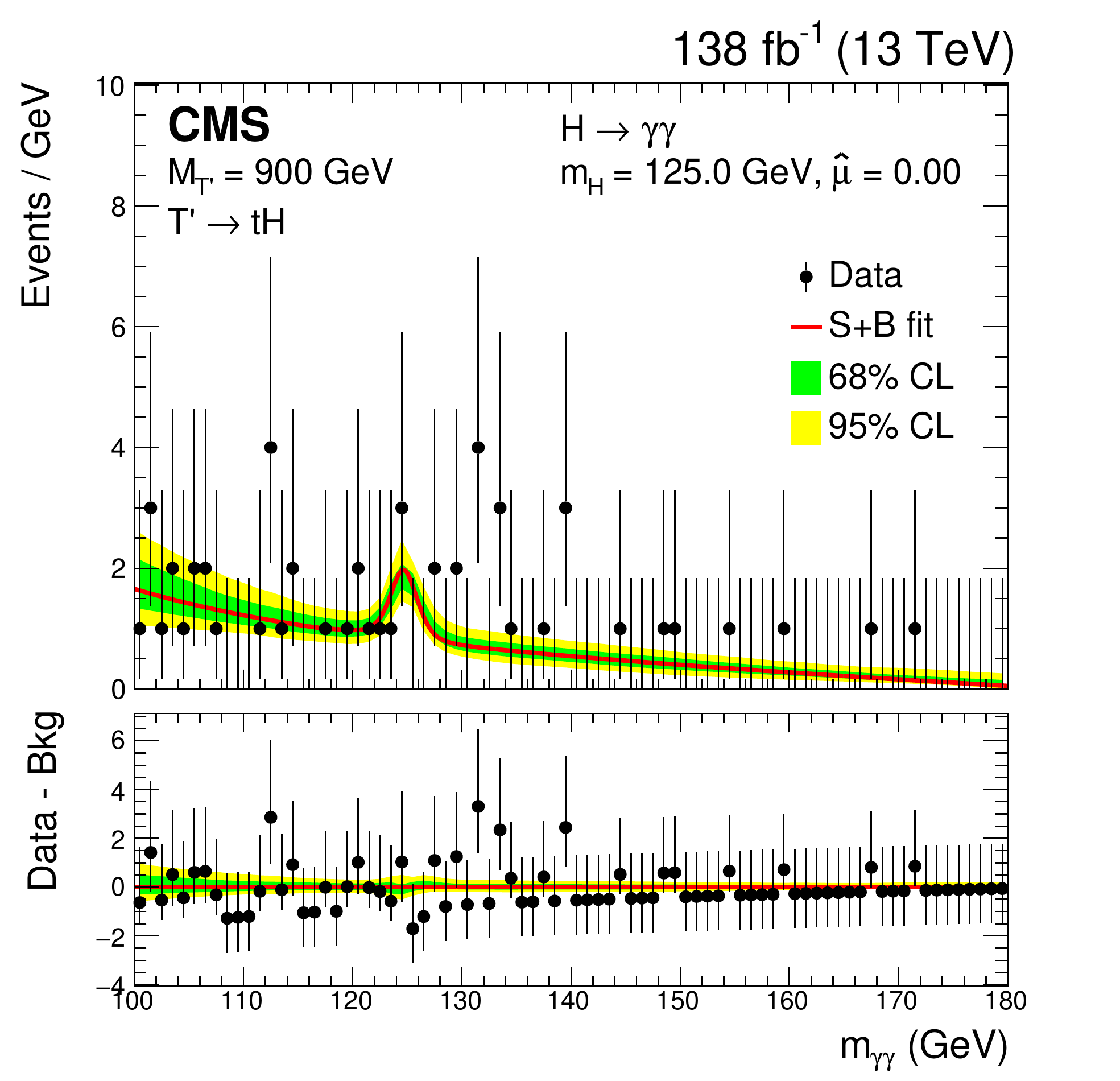}
    \includegraphics[width=0.49\textwidth]{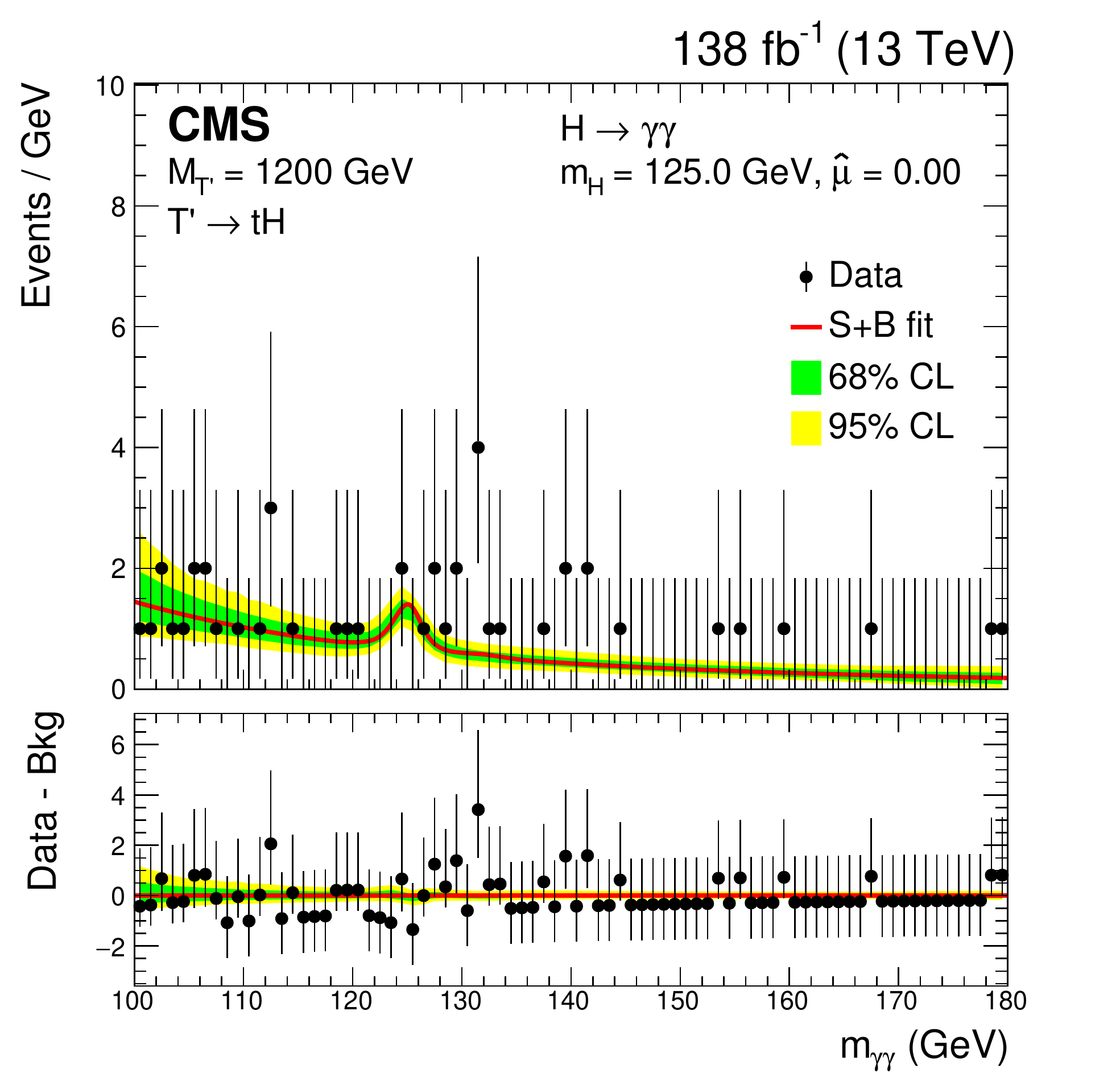}
    \caption{The combined, leptonic plus hadronic, distributions for data (black dots) and \mgg
        signal-plus-background model fits (red line) for a VLQ signal with \MTprime of 600 (upper left), 900 (upper right),
        and 1200\GeV (lower). The green (yellow) band represents the 68\% (95\%) \CL in the background
        component of the fit. The peak in the background component
        shows the considered irreducible SM Higgs boson (ggH, VBF, VH, \ttH and tH) contribution. Here, $\hat{\mu}$ is the best fit
        value of the signal strength parameter $\mu$, which is zero for the three \MTprime values considered.
        The lower panel shows the residuals after the subtraction of the background component.}
    \label{fig:result_SplusB_and_2D_mass_plots}
\end{figure}

\section{Results}
The combined (leptonic and hadronic analyses) data distributions and the corresponding signal-plus-background
model fits of the \mgg distribution are shown in Fig.~\ref{fig:result_SplusB_and_2D_mass_plots} for
\MTprime values of 600, 900, and 1200\GeV.
No statistically significant excesses above the SM backgrounds in any of the
channels or mass ranges are observed. Upper limits on the signal strength modifiers
$\mu_{\mathrm{obs}} = (\sigma)_{\mathrm{obs}}/(\sigma)_{\mathrm{th}}$ and $\mu_{\mathrm{exp}} = (\sigma)_{\mathrm{exp}}/(\sigma)_{\mathrm{th}}$, are derived for different
\MTprime, using a maximum likelihood fit of the \mgg distributions, keeping the $M_{\PH}$ parameter
of the model fixed at 125\GeV.
The expected and observed upper limits are estimated at the 95\% \CL based on the \CLs criterion~\cite{Junk:1999kv, ALRead_2002} using
the asymptotic approximation~\cite{asymptoticlimit, Feldman:1997qc} for the test statistic.
The results are verified with pseudoexperiments
and are cross-checked with detailed bias studies on the parameter $\mu$.

\begin{figure}[htbp!]
    \centering
    \includegraphics[width=0.70\textwidth]{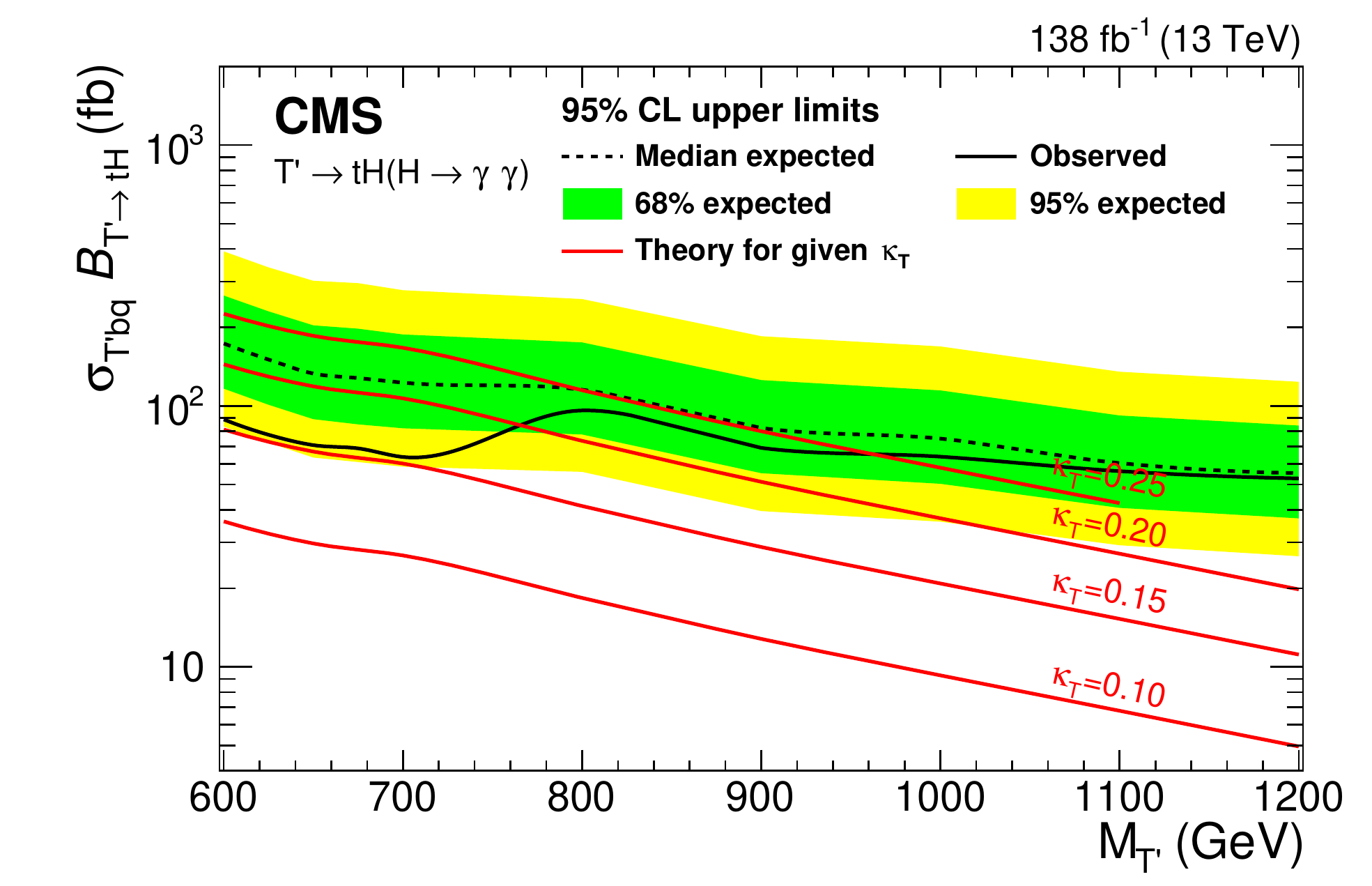}
    \caption{The combined, leptonic plus hadronic, expected (dotted black) and observed (solid black) upper limits at 95\%
        \CL on $\sigma_{\Tprime\bq}\mathcal{B_{\TtH}}$ are displayed as a function of \MTprime. The green (yellow)
        band represents the 68\% (95\%) of the limit values expected under the background-only
        hypothesis. The theoretical cross sections for the singlet \Tprime production with
        representative $\kappaT$-values fixed at 0.1, 0.15, 0.2 and 0.25 (for $\Gamma/\MTprime < 5\%$)
        are shown as red lines.}
    \label{fig:result_xsec_combined}
\end{figure}

\begin{figure}[htbp!]
    \centering
    \includegraphics[width=0.70\textwidth]{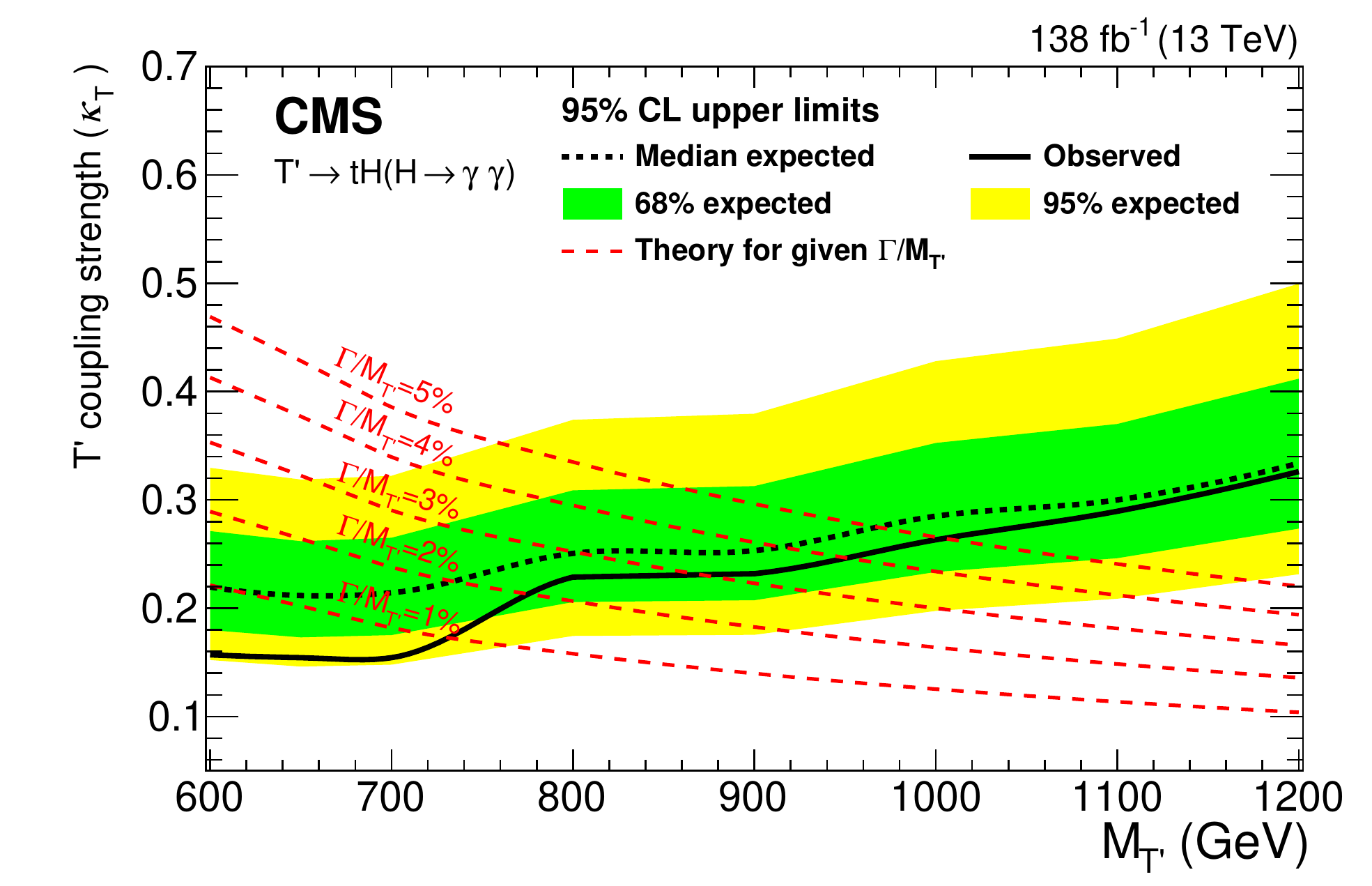}
    \caption{The combined, leptonic plus hadronic, expected (dotted black) and observed (solid black) upper limits at
        95\% \CL on the \Tprime coupling to third-generation quarks, $\kappaT$,
        under the narrow width approximation displayed as a function of \MTprime.
        The green (yellow) band represents the 68\% (95\%) of the limit values expected under the
        background-only hypothesis. The theoretical $\kappaT$ values corresponding to the
        $\Gamma/\MTprime$-values fixed at 1, 2, 3, 4, and 5\% are shown as red dashed lines.}
    \label{fig:result_kappa}
\end{figure}

Finally, the upper limits on $\mu_{\mathrm{obs}}$ and $\mu_{\mathrm{exp}}$ are translated into the
upper limits on $\sigma_{\Tprime\bq}\mathcal{B}_{\TtH}$, as displayed in Fig.~\ref{fig:result_xsec_combined}
together with the theoretical cross sections for the singlet \Tprime production with
representative $\kappaT$-values fixed at 0.1, 0.15, 0.2 and 0.25 (for $\Gamma/\MTprime < 5\%$).
Similarly, the upper limits on the coupling parameter of \Tprime with the SM particles (\kappaT)
under the narrow width approximation is displayed in Fig.~\ref{fig:result_kappa} with theoretical
\kappaT-values corresponding to the $\Gamma/\MTprime$-values fixed at 1, 2, 3, 4, and 5\%.
Tabulated results are provided in the HEPData record for this analysis~\cite{hepdata}. The limits
obtained separately in the leptonic and hadronic channels are available in the HEPData record,
where it can be seen that the hadronic channel dominates the sensitivity.

\section{Summary}
A search for a vector-like quark decaying to a top quark and a Higgs boson that decays into two photons, $\Tprime \to \tH$ (\Hgg), has been performed
using proton-proton collision data at $\sqrt{s} = 13\TeV$ recorded with the CMS detector
in 2016--2018, and corresponding to an integrated luminosity of 138\fbinv.
The search has been carried out based on a model of \Tprime electroweak production in a narrow
width approximation with a ratio of \Tprime width relative to its mass ($\Gamma/\MTprime$) $\approx 1\%$.
The sensitivity of this analysis extends up to
$\Gamma/\MTprime\approx 5\%$, which roughly corresponds to the experimental resolution of \MTprime.
Both the hadronic and leptonic decay modes of the top quark are considered in the search.
A novel multivariate analysis incorporating three separately optimized boosted decision trees is exploited to separate likely signal
events from background processes, including the standard model production of Higgs bosons.
No statistically significant excess over the expected background prediction is observed.
Assuming a coupling to third generation quarks of $\kappaT=0.25$ and a relative decay width of
$\Gamma/\MTprime < 5\%$, the electroweak production of a singlet \Tprime quark is excluded up to a mass of 960\GeV at 95\% confidence level.
This search for a vector-like quark, \Tprime, is the most sensitive to date for \MTprime up to 1.1\TeV, among searches exploring the same production mechanism.

\begin{acknowledgments}
We congratulate our colleagues in the CERN accelerator departments for the excellent performance of the LHC and thank the technical and administrative staffs at CERN and at other CMS institutes for their contributions to the success of the CMS effort. In addition, we gratefully acknowledge the computing centers and personnel of the Worldwide LHC Computing Grid and other centers for delivering so effectively the computing infrastructure essential to our analyses. Finally, we acknowledge the enduring support for the construction and operation of the LHC, the CMS detector, and the supporting computing infrastructure provided by the following funding agencies: BMBWF and FWF (Austria); FNRS and FWO (Belgium); CNPq, CAPES, FAPERJ, FAPERGS, and FAPESP (Brazil); MES and BNSF (Bulgaria); CERN; CAS, MoST, and NSFC (China); MINCIENCIAS (Colombia); MSES and CSF (Croatia); RIF (Cyprus); SENESCYT (Ecuador); MoER, ERC PUT and ERDF (Estonia); Academy of Finland, MEC, and HIP (Finland); CEA and CNRS/IN2P3 (France); BMBF, DFG, and HGF (Germany); GSRI (Greece); NKFIH (Hungary); DAE and DST (India); IPM (Iran); SFI (Ireland); INFN (Italy); MSIP and NRF (Republic of Korea); MES (Latvia); LAS (Lithuania); MOE and UM (Malaysia); BUAP, CINVESTAV, CONACYT, LNS, SEP, and UASLP-FAI (Mexico); MOS (Montenegro); MBIE (New Zealand); PAEC (Pakistan); MES and NSC (Poland); FCT (Portugal); MESTD (Serbia); MCIN/AEI and PCTI (Spain); MOSTR (Sri Lanka); Swiss Funding Agencies (Switzerland); MST (Taipei); MHESI and NSTDA (Thailand); TUBITAK and TENMAK (Turkey); NASU (Ukraine); STFC (United Kingdom); DOE and NSF (USA).
    
\hyphenation{Rachada-pisek} Individuals have received support from the Marie-Curie program and the European Research Council and Horizon 2020 Grant, contract Nos.\ 675440, 724704, 752730, 758316, 765710, 824093, 884104, and COST Action CA16108 (European Union); the Leventis Foundation; the Alfred P.\ Sloan Foundation; the Alexander von Humboldt Foundation; the Belgian Federal Science Policy Office; the Fonds pour la Formation \`a la Recherche dans l'Industrie et dans l'Agriculture (FRIA-Belgium); the Agentschap voor Innovatie door Wetenschap en Technologie (IWT-Belgium); the F.R.S.-FNRS and FWO (Belgium) under the ``Excellence of Science -- EOS" -- be.h project n.\ 30820817; the Beijing Municipal Science \& Technology Commission, No. Z191100007219010; the Ministry of Education, Youth and Sports (MEYS) of the Czech Republic; the Hellenic Foundation for Research and Innovation (HFRI), Project Number 2288 (Greece); the Deutsche Forschungsgemeinschaft (DFG), under Germany's Excellence Strategy -- EXC 2121 ``Quantum Universe" -- 390833306, and under project number 400140256 - GRK2497; the Hungarian Academy of Sciences, the New National Excellence Program - \'UNKP, the NKFIH research grants K 124845, K 124850, K 128713, K 128786, K 129058, K 131991, K 133046, K 138136, K 143460, K 143477, 2020-2.2.1-ED-2021-00181, and TKP2021-NKTA-64 (Hungary); the Council of Science and Industrial Research, India; the Latvian Council of Science; the Ministry of Education and Science, project no. 2022/WK/14, and the National Science Center, contracts Opus 2021/41/B/ST2/01369 and 2021/43/B/ST2/01552 (Poland); the Funda\c{c}\~ao para a Ci\^encia e a Tecnologia, grant CEECIND/01334/2018 (Portugal); the National Priorities Research Program by Qatar National Research Fund; MCIN/AEI/10.13039/501100011033, ERDF ``a way of making Europe", and the Programa Estatal de Fomento de la Investigaci{\'o}n Cient{\'i}fica y T{\'e}cnica de Excelencia Mar\'{\i}a de Maeztu, grant MDM-2017-0765 and Programa Severo Ochoa del Principado de Asturias (Spain); the Chulalongkorn Academic into Its 2nd Century Project Advancement Project, and the National Science, Research and Innovation Fund via the Program Management Unit for Human Resources \& Institutional Development, Research and Innovation, grant B05F650021 (Thailand); the Kavli Foundation; the Nvidia Corporation; the SuperMicro Corporation; the Welch Foundation, contract C-1845; and the Weston Havens Foundation (USA).    
\end{acknowledgments}

\bibliography{auto_generated}
\cleardoublepage \appendix\section{The CMS Collaboration \label{app:collab}}\begin{sloppypar}\hyphenpenalty=5000\widowpenalty=500\clubpenalty=5000
\cmsinstitute{Yerevan Physics Institute, Yerevan, Armenia}
{\tolerance=6000
A.~Tumasyan\cmsAuthorMark{1}\cmsorcid{0009-0000-0684-6742}
\par}
\cmsinstitute{Institut f\"{u}r Hochenergiephysik, Vienna, Austria}
{\tolerance=6000
W.~Adam\cmsorcid{0000-0001-9099-4341}, J.W.~Andrejkovic, T.~Bergauer\cmsorcid{0000-0002-5786-0293}, S.~Chatterjee\cmsorcid{0000-0003-2660-0349}, K.~Damanakis\cmsorcid{0000-0001-5389-2872}, M.~Dragicevic\cmsorcid{0000-0003-1967-6783}, A.~Escalante~Del~Valle\cmsorcid{0000-0002-9702-6359}, P.S.~Hussain\cmsorcid{0000-0002-4825-5278}, M.~Jeitler\cmsAuthorMark{2}\cmsorcid{0000-0002-5141-9560}, N.~Krammer\cmsorcid{0000-0002-0548-0985}, L.~Lechner\cmsorcid{0000-0002-3065-1141}, D.~Liko\cmsorcid{0000-0002-3380-473X}, I.~Mikulec\cmsorcid{0000-0003-0385-2746}, P.~Paulitsch, F.M.~Pitters, J.~Schieck\cmsAuthorMark{2}\cmsorcid{0000-0002-1058-8093}, R.~Sch\"{o}fbeck\cmsorcid{0000-0002-2332-8784}, D.~Schwarz\cmsorcid{0000-0002-3821-7331}, M.~Sonawane\cmsorcid{0000-0003-0510-7010}, S.~Templ\cmsorcid{0000-0003-3137-5692}, W.~Waltenberger\cmsorcid{0000-0002-6215-7228}, C.-E.~Wulz\cmsAuthorMark{2}\cmsorcid{0000-0001-9226-5812}
% ...
\par}
\vskip\cmsinstskip
\dag:~Deceased\\
$^{1}$Also at Yerevan State University, Yerevan, Armenia\\
$^{2}$Also at TU Wien, Vienna, Austria\\
% ...
\end{sloppypar}
\end{document}